\documentstyle{mn}
\input{psfig}
\newcommand{\etal}{{et al.~}}

\newcommand{\kmsmpc}{\>{\rm km}\,{\rm s}^{-1}\,{\rm Mpc}^{-1}}

\newcommand{\Mpc}{\>{\rm Mpc}}

\newcommand{\Msun}{\>{\rm M_{\odot}}}

\newcommand{\yr}{\>{\rm yr}}

\newcommand{\beq}{\begin{equation}}
\newcommand{\eeq}{\end{equation}}

\def\gtsima{$\; \buildrel > \over \sim \;$}
\def\ltsima{$\; \buildrel < \over \sim \;$}
\def\prosima{$\; \buildrel \propto \over \sim \;$}
\def\gsim{\lower.7ex\hbox{\gtsima}}
\def\lsim{\lower.7ex\hbox{\ltsima}}
\def\simgt{\lower.7ex\hbox{\gtsima}}
\def\simlt{\lower.7ex\hbox{\ltsima}}
\def\simpr{\lower.7ex\hbox{\prosima}}
\def\la{\lsim}
\def\ga{\gsim}
\def\lta{\la}
\def\gta{\ga}

\newcommand{\apj}{ApJ}
\newcommand{\apjs}{ApJS}
\newcommand{\aj}{AJ}
\newcommand{\mnras}{MNRAS}
\newcommand{\aap}{A\&A}
\newcommand{\aaps}{A\&AS}
\newcommand{\araa}{ARA\&A}
\newcommand{\nat}{Nature}

\newdimen\hssize
\newdimen\hdsize
\begin{document}


\title[On the Origin of the Dichotomy of Early-Type Galaxies]
      {On the Origin of the Dichotomy of Early-Type Galaxies:
       The Role of Dry Mergers and AGN Feedback}
\author[Kang et al.]
       {X. Kang$^{1,2}$\thanks{E-mail:kang@mpia.de}, 
        Frank C. van den Bosch$^{1}$, A. Pasquali$^{1}$\\
        $^1$Max-Planck-Institute for Astronomy, K\"onigstuhl 17, D-69117
            Heidelberg, Germany\\
        $^2$Shanghai Astronomical Observatory; the Partner Group of MPA,
           Nandan Road 80, Shanghai 200030, China}


\date{}

\pagerange{\pageref{firstpage}--\pageref{lastpage}}
\pubyear{2000}

\maketitle

\label{firstpage}


\begin{abstract}
  Using a semi-analytical model for  galaxy formation, combined with a
  large  $N$-body simulation,  we   investigate   the origin  of   the
  dichotomy among  early-type galaxies.  In qualitative agreement with
  previous studies and  with numerical simulations,  we find that boxy
  galaxies originate from mergers with a progenitor mass ratio $n < 2$
  and with a combined cold  gas  mass fraction  $F_{\rm cold} <  0.1$.
  Our   model  accurately reproduces  the observed   fraction  of boxy
  systems as a function of luminosity and  halo mass, for both central
  galaxies and satellites.     After correcting for  the  stellar mass
  dependence, the  properties of the last  major merger  of early-type
  galaxies are   independent   of their   halo  mass.  This   provides
  theoretical support for the conjecture of Pasquali \etal (2007) that
  the stellar mass (or luminosity) of an early-type galaxy is the main
  parameter that governs its isophotal shape.   If wet and dry mergers
  mainly  produce disky   and   boxy early-types, respectively,    the
  observed dichotomy of  early-type galaxies has a natural explanation
  within the hierarchical framework  of structure formation.  Contrary
  to naive expectations, the dichotomy is independent of AGN feedback.
  Rather, we argue that it owes to the fact  that more massive systems
  (i)  have more massive  progenitors, (ii)  assemble later, and (iii)
  have a larger  fraction  of early-type  progenitors.  Each of  these
  three trends causes the cold gas mass fraction of the progenitors of
  more  massive early-types to  be  lower, so  that  their last  major
  merger involved  less  cold gas (was   more  ``dry'').  Finally, our
  model predicts  that (i)  less than  10  percent of  all  early-type
  galaxies form in   major    mergers that  involve   two   early-type
  progenitors,  (ii) more   than 95 percent   of all  boxy  early-type
  galaxies with $M_* \lta 2 \times 10^{10} h^{-1} \Msun$ are satellite
  galaxies, and (iii) about 70 percent of all  low mass early-types do
  not  form  a  supermassive black hole  binary   at their  last major
  merger.  The latter  may help to   explain why low  mass early-types
  have central cusps, while their massive counterparts have cores.
\end{abstract}


\begin{keywords}
dark matter  --- 
galaxies: elliptical and lenticular --- 
galaxies: interactions ---
galaxies: structure ---
galaxies: formation
\end{keywords}


\section{Introduction}
\label{sec:intro}

Ever since the  seminal paper by Davies \etal (1983)  it is clear that
early-type galaxies (ellipticals  and lenticulars, hereafter ETGs) can
be split in two distinct sub-classes.  Davies \etal showed that bright
ETGs typically  have little rotation, such that  their flattening must
originate from  anisotropic pressure.  This is  consistent with bright
ellipticals being in general triaxial.  Low luminosity ellipticals, on
the other hand, typically have rotation velocities that are consistent
with them being oblate isotropic rotators (see Emsellem \etal 2007 and
Cappellari  \etal 2007  for a  more contemporary  description).  These
different  kinematic  classes  also  have different  morphologies  and
different central surface  brightness profiles.  In particular, bright,
pressure-supported systems usually  have boxy isophotes and luminosity
profiles that break from steep outer power-laws to shallow inner cusps
(often  called  `cores').   The  low  luminosity,  rotation  supported
systems,  on  the  other   hand,  often  reveal  disky  isophotes  and
luminosity  profiles with  a steep  central cusp  (e.g.,  Bender 1988;
Nieto  \etal 1988;  Ferrarese \etal  1994; Gebhardt  \etal  1996; Rest
\etal 2001; Lauer  \etal 2005, 2006). Finally, the  bimodality of ETGs
has also  been found to  extend to their  radio and X-ray  properties. 
Objects  which are  radio-loud and/or  bright in  soft  X-ray emission
generally have boxy isophotes, while disky ETGs are mostly radio-quiet
and faint  in soft X-rays  (Bender \etal 1989; Pellegrini  1999, 2005;
Ravindranath \etal 2001).

Recently, Hao  \etal (2006) constructed  a homogeneous samples  of 847
ETGs from  the SDSS DR4  catalogue (Adelman-McCarthy \etal  2006), and
analyzed  their isophotal  shapes. This  sample was  used  by Pasquali
\etal (2007:  hereafter P07) to investigate the  relative fractions of
disky  and boxy  ETGs  as  function of  luminosity,  stellar mass  and
environment.  They  found that  the disky fraction  decreases smoothly
with increasing  ($B$-band) luminosity,  stellar mass, and  halo mass,
where the latter is obtained from the SDSS group catalogue of Weinmann
\etal (2006).  In addition, the disky  fraction is found  to be higher
for satellite galaxies than for central galaxies in a halo of the same
mass.  These data  provide a powerful benchmark against  which to test
models for the formation of ETGs.

Within the  framework of hierarchical  structure formation, elliptical
galaxies are  generally assumed to  form through major  mergers (e.g.,
Toomre \&  Toomre 1972; Schweizer  1982; Barnes 1988;  Hernquist 1992;
Kauffmann, White \& Guiderdoni 1993).   In this case, it seems logical
that the bimodality in their isophotal and kinematical properties must
somehow be  related to the  details of their merger  histories.  Using
dissipationless $N$-body simulations it has been shown that equal mass
mergers of  disk galaxies mainly  result in slowly rotating  ETGs with
boxy (but sometimes disky) isophotes,  while mergers in which the mass
ratio  between the  progenitor disks  is significantly  different from
unity mainly yields disky ETGs (Negroponte \& White 1983; Barnes 1988;
Hernquist 1992; Bendo  \& Barnes 2000; Naab \&  Burkert 2003; Bournaud
\etal 2004,  2005; Naab \& Trujillo 2006).   However, simulations that
also include a dissipative gas component and star formation have shown
that the presence of even a relatively small amount of cold gas in the
progenitors  results  in  a  merger  remnant that  resembles  a  disky
elliptical even  when the  mass ratio of  the progenitors is  close to
unity (Barnes \&  Hernquist 1996; Naab \etal 2006a;  Cox \etal 2006a). 
This suggests that  boxy ETGs can only form out  of dry, major mergers
(see also discussion in Faber \etal 1997).  In this paper we test this
paradigm using  a semi-analytical model  for galaxy formation  and the
observational constraints of P07.

Our study is similar to those of Khochfar  \& Burkert (2005; hereafter
KB05)  and  Naab   \etal   (2006b;  hereafter  N06),    who  also used
semi-analytical models to explore  whether the dichotomy of elliptical
galaxies   can  be related  to  their  merger properties. However, our
analysis differs from theirs on the following grounds.
\begin{itemize}
  
\item We use  a numerical $N$-body simulation to  construct the merger
  histories of dark matter haloes.  The models of KB05 and N06, on the
  other hand, used merger  trees based on the extended Press-Schechter
  (EPS) formalism (e.g.,  Lacey \& Cole 1993).  As  we will show, this
  results in significant differences.
  
\item Because of our use of  numerical $N$-body simulations, our model
  more  accurately   traces the  dynamical   evolution of  dark matter
  subhaloes with their associated  satellite galaxies. In  particular,
  it  takes proper account of  dynamical friction, tidal stripping and
  the merging between subhaloes.
  
\item  Contrary to KB05  and N06,  we include  a prescription  for the
  feedback from  active galactic  nuclei (AGN) in  our semi-analytical
  model. 
  
\item Our  semi-analytical model is tuned to  reproduce the luminosity
  function  and  the  color-bimodality  of the  redshift  zero  galaxy
  population (see Kang \etal 2005).  The  works of KB05 and N06 do not
  mention such a comparison.
 
\item Our criteria for the  production of boxy ETGs are different from
  those used in KB05 and N06.

\item We use a much larger, more homogeneous data set to constrain the
  models.

\end{itemize}

This  paper is  organized as  follows. In  Section~\ref{sec:method} we
describe our N-body simulation  and semi-analytical model, and outline
the     methodology.     The      results     are     described     in
Section~\ref{sec:results} and  discussed in Section~\ref{sec:disc}. We
summarize our findings in Section~\ref{sec:concl}

\section{Methodology}
\label{sec:method}

The aim of this paper is to investigate to what extend semi-analytical
models  of galaxy  formation can  reproduce  the ecology  of ETGs,  in
particular  the fractions  of disky  and boxy  systems as  function of
luminosity and halo mass. Partially motivated by numerical simulations
of galaxy mergers, both with and  without gas, we adopt a framework in
which (i) ETGs  are red and dominated by  a spheroidal component, (ii)
ETGs are  the outcome of major  mergers, (iii) the remnant  is boxy if
the merger is sufficiently  ``dry'' (i.e., the progenitors have little
or no cold gas) and sufficiently ``major'' (i.e., the progenitors have
roughly  equal masses)  and (iv)  a  boxy elliptical  becomes a  disky
elliptical  if newly  accreted  material builds  a sufficiently  large
stellar disk.
                                
\subsection{N-body simulation and model descriptions}
\label{sec:model}

In order to have accurate merger  trees, and to be  able to follow the
dynamical evolution of     satellite galaxies,  we   use  a  numerical
simulation  of the evolution   of dark matter  which we  populate with
galaxies using   a state-of-the-art semi-analytical  model for  galaxy
formation.  The numerical  simulation has been carried  out by Jing \&
Suto (2002) using a vectorized-parallel  P$^3$M code.  It follows  the
evolution of $512^{3}$ particles in a  cosmological box of $100 h^{-1}
\Mpc$,  assuming  a flat   $\Lambda$CDM  `concordance'  cosmology with
$\Omega_m =0.3$,  $\sigma_{8}=0.9$,  and $h = H_0/100\kmsmpc   = 0.7$.
Each particle has a  mass of $6.2 \times 10^{8}  h^{-1}  \Msun$.  Dark
matter haloes   are identified   using the   friends-of-friends  (FOF)
algorithm with a linking length equal to $0.2$ times the mean particle
separation.  For each  halo   thus identified we compute    the virial
radius, $r_{\rm vir}$, defined  as the {\it spherical} radius centered
on  the most  bound particle inside   of which the  average density is
$340$ times the average density of the  Universe (cf.  Bryan \& Norman
1998). The virial mass is simply defined  as the mass of all particles
that have halocentric   radii $r \leq   r_{\rm vir}$.  Since  our  FOF
haloes  have a characteristic  overdensity of $\sim  180$ (e.g., White
2002), the virial mass is typically smaller than the FOF mass.

Dark matter subhaloes within   each FOF (parent) halo  are  identified
using  the SUBFIND routine described  in Springel \etal (2001). In the
present study,  we use all  haloes and subhaloes  with masses  down to
$6.2   \times   10^{9}h^{-1}M_{\odot}$   (10  particles).   Using   60
simulation  outputs between   $z=15$  and $z=0$,   equally   spaced in
$\log(1+z)$,  Kang \etal (2005;  hereafter K05) constructed the merger
history for each (sub)halo in the  simulation box, which are then used
in the semi-analytical model.  In what follows, whenever we refer to a
{\it halo}, we mean a  virialized object which  is not a sub-structure
of a  larger virialized object, while   {\it subhaloes} are virialized
objects   that orbit within a   halo.  In  addition, (model)  galaxies
associated with  haloes  and  subhaloes  are  referred  to as  central
galaxies and satellites, respectively.

The semi-analytical  model used to  populate the  haloes and subhaloes
with galaxies is described  in detail in K05,  to  which we refer  the
reader for   details.  Briefly, the  model  assumes  that the baryonic
material  accreted  by a  dark  matter halo  is  heated  to the virial
temperature.  The gas then cools  radiatively and settles down into  a
centrifugally  supported  disk, in which the  star   formation rate is
proportional to  the   total   amount  of  cold gas,    and  inversely
proportional to the dynamical time  of  the disk. The energy  feedback
from supernova  is related to  the initial stellar mass function (IMF)
and proportional to the  star formation rate.  It  is assumed that the
gas that is reheated  by supernova feedback  remains bound to the host
halo so that it can cool back onto the disk at later stages.  When the
subhalo  associated with  a    satellite galaxy is  dissolved   in the
numerical  simulation  the   satellite galaxy becomes  an   ``orphan''
galaxy,  which is  assumed to merge  with   the central galaxy  of the
parent halo after a    dynamical  friction time  (computed    assuming
standard Chandrasekhar dynamical  friction).  When two galaxies merge,
the outcome  is  assumed  to depend on  their   mass ratio  $n  \equiv
M_1/M_2$ with $M_1 \geq M_2$.  If $n \leq  3$ the merger is assumed to
result  in the formation  of an elliptical galaxy,  and we  speak of a
``major merger''.  Any   cold gas  available  in both  progenitors  is
turned into stars.  This  is supported by  hydrodynamical simulations,
which show that major mergers cause the cold gas to flow to the center
where the  resulting   high gas density  triggers  a  starburst (e.g.,
Barnes \&  Hernquist  1991, 1996;  Mihos \&  Hernquist  1996; Springel
2000; Cox  \etal 2006a,b; Di Matteo  \etal 2007).  A  new disk of cold
gas  and stars may  form around the elliptical  if new gas can cool in
the halo  of the merger  remnant, thus  giving  rise to  a  disk-bulge
system.  If $n > 3$ we speak  of a ``minor  merger'' and we simply add
the cold gas of the less massive progenitor to that of the disk of the
more massive progenitor, while its stellar  mass is added to the bulge
of the massive progenitor.  The  semi-analytical model also includes a
prescription  for ``radio-mode'' AGN  feedback  as described  in Kang,
Jing \& Silk (2006; see  also Section~\ref{sec:AGN}).  This ingredient
is essential to  prevent significant amounts of  star formation in the
brightest galaxies,   and thus  to  ensure  that  these   systems  are
predominantly members of the  red sequence (e.g., Cattaneo \etal 2006;
De Lucia  \etal 2006; Bower \etal 2006;  Croton \etal 2006).  Finally,
luminosities  for all model galaxies  are computed using the predicted
star formation histories and  the stellar population models of Bruzual
\& Charlot   (2003).  Throughout  we  assume a  Salpeter  IMF  and  we
self-consistently model the metalicities  of gas and stars, including
metal-cooling.

As  shown in K05  and Kang  \etal (2006)  this model  accurately fits,
among  others, the  galaxy  luminosity function  at  $z=0$, the  color
bimodality of the  $z=0$ galaxy population, and the  number density of
massive, red  galaxies out to  $z \sim 3$.  We emphasize that  in this
paper we use this model without changing any of its parameters.
\begin{figure*}
\centerline{\psfig{figure=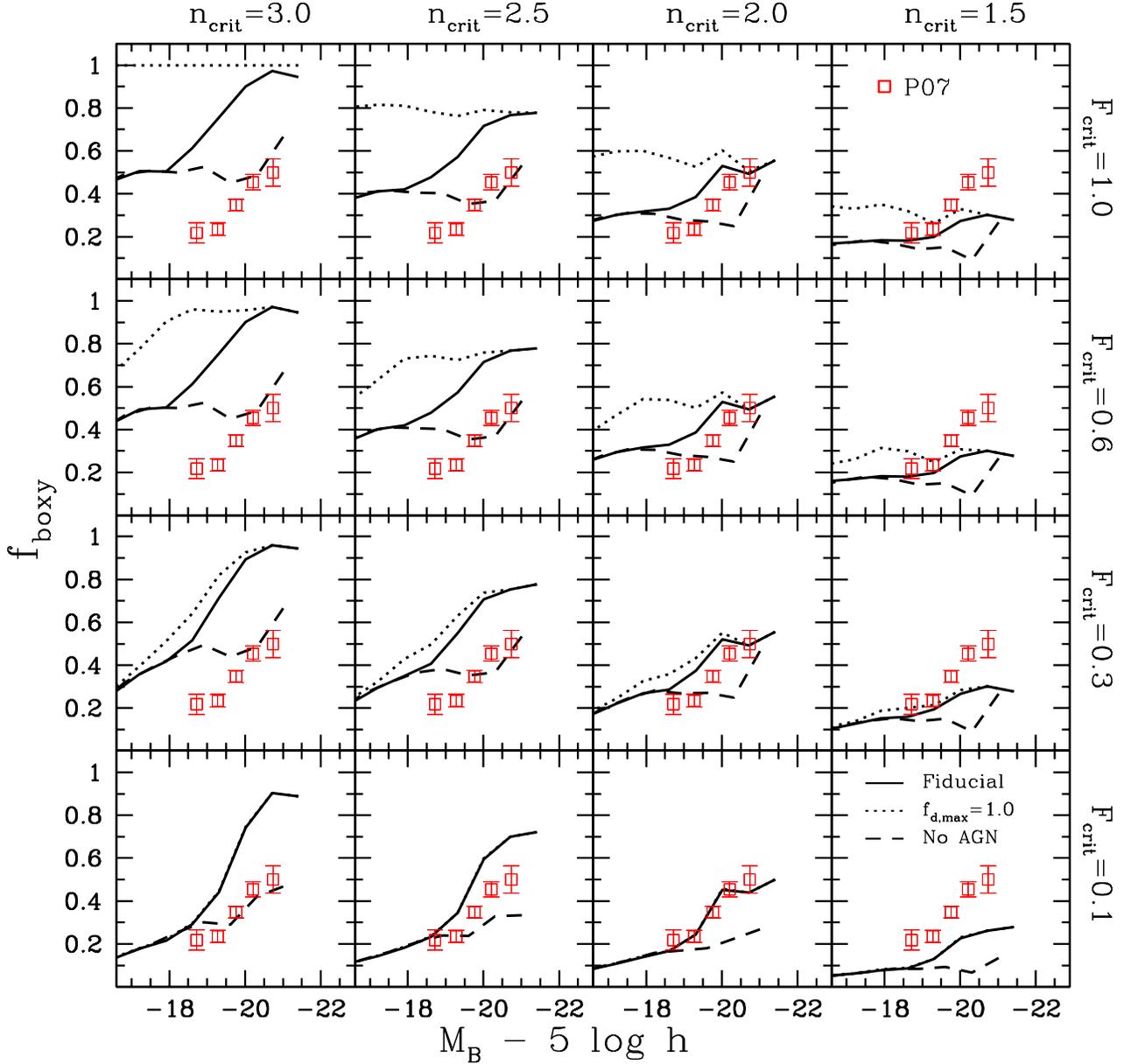,width=0.95\hdsize}}
\caption{The boxy  fraction of ETGs as function of their $B$-band 
  magnitude  (in   the  AB  system).   The  open,   red  squares  with
  (Poissonian)  errorbars  correspond to  the  data  of  P07, and  are
  duplicated  in  each  panel.  The  solid, dotted  and  dashed  lines
  correspond  to the three  models discussed  in the  text.  Different
  columns  and rows correspond  to different  values for  the critical
  progenitor  mass ratio, $n_{\rm  crit}$, and  the critical  cold gas
  mass fraction, $F_{\rm crit}$, respectively, as indicated.}
\label{fig:fboxy}
\end{figure*}

\subsection{Predicting Isophotal Shapes}
\label{sec:isoph}

In our  model we  determine whether an  elliptical galaxy is  disky or
boxy  as follows.  Using  the output  at $z=0$  we first  identify the
early-type  (E/S0) galaxies  based  on two  criteria.   First of  all,
following Simien \& de Vaucouleurs (1986), we demand that an ETG has a
bulge-to-disk  ratio in the  $B$-band of  $L_{B,{\rm bulge}}/L_{B,{\rm
    total}} \geq 0.4$. In addition,  we require the $B-V$ color of the
galaxy to be red.  Following Hao \etal (2006) and P07, we adopt $B-V >
0.8$. We  have verified that none  of our results are  sensitive to the
exact choice of these  selection criteria.  

Having thus identified all ETGs  at $z=0$, we subsequently trace their
formation histories  back until their last major  merger, and register
the mass ratio $n$ of the merger  event, as well as the total cold gas
mass fraction at that epoch, defined as
\begin{equation}
\label{fcolddef}
F_{\rm cold} \equiv {\sum\limits_{i=1}^2 M_{{\rm cold},i} \over 
\sum\limits_{i=1}^2 (M_{{\rm cold},i} + M_{*,i})}
\end{equation}
Here  $M_{{\rm cold},i}$  and  $M_{*,i}$  are the  cold  gas mass  and
stellar mass of progenitor $i$, respectively.  We adopt the hypothesis
that the  merger results in  a boxy elliptical  if, and only if,  $n <
n_{\rm crit}$ and $F_{\rm cold} < F_{\rm crit}$.  The main aim of this
paper is  to use the  data of P07  to constrain the values  of $n_{\rm
  crit}$ and $F_{\rm crit}$, and to investigate whether a model can be
found that  is consistent with  the observed fraction of  boxy (disky)
ETGs as function of both galaxy luminosity and halo mass.

The final ingredient for determining  whether an ETG  is disky or boxy
is the potential regrowth  of a stellar disk.   Between its last major
merger and the present  day, new gas in  the  halo of the  remnant may
cool out to form a new disk. In addition, the ETG may also accrete new
stars and cold  gas via minor mergers (those  with $n > 3$).  Any cold
gas in those  accreted systems is  added to the  new disk, where it is
allowed to form  new  stars.  Whenever  the   stellar disk has   grown
sufficiently     massive,  its presence will     reveal  itself in the
isophotes, and the system changes from being boxy  to being disky.  To
take this effect into account, we follow KB05 and we reclassify a boxy
system as disky if  at $z=0$ it  has grown a  disk with a stellar mass
that contributes more  than a fraction  $f_{d,{\rm max}}$ of the total
stellar mass in the  galaxy.  In our fiducial  model we set $f_{d,{\rm
    max}} = 0.2$.  This is motivated by  Rix \& White (1990), who have
shown that if  an embedded stellar  disk contains more  than $\sim 20$
percent of the  total stellar mass, the isophotes  of its  host galaxy
become disky.  Note that the same value for $f_{d,{\rm max}}$ has also
been used in the analysis of KB05.
\begin{figure*}
\centerline{\psfig{figure=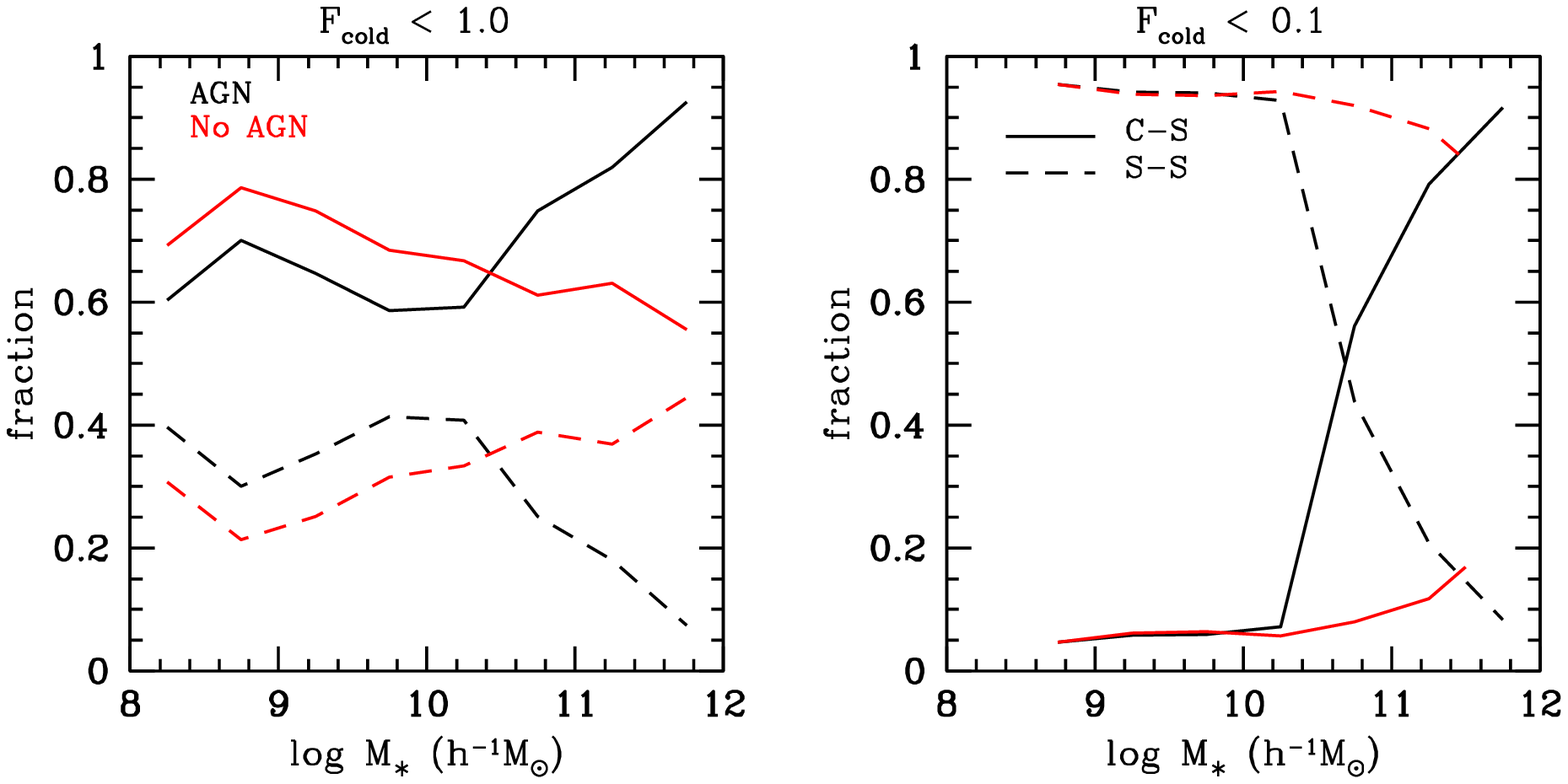,width=0.95\hdsize}}
\caption{The fractions  of  last  major mergers between   centrals and
  satellites  (C-S; solid  lines)  and  between  two satellites  (S-S;
  dashed lines) that result in the formation  of an ETG as function of
  its stellar mass at $z=0$. Results are shown for all ETGs (left-hand
  panel) and for those   with $F_{\rm cold}<0.1$  (right-hand  panel).
  Black and    red lines correspond to  models   with and  without AGN
  feedback, respectively. Low mass ETGs that form  in dry mergers, and
  hence end up  being boxy, mainly  form out  of S-S mergers.   At the
  massive end, the fraction of ETGs that form  out of S-S mergers with
  $F_{\rm  cold}=0.1$ depends strongly on   the presence or absence of
  AGN feedback. See text for detailed discussion.}
\label{fig:cs}
\end{figure*}

\section{Results}
\label{sec:results}

In Figure~\ref{fig:fboxy}  we show the fraction of  boxy ETGs, $f_{\rm
  boxy}$, as a  function of the luminosity in the  $B$-band (in the AB
system).  Open squares  with errorbars (reflecting Poisson statistics)
correspond to  the data of P07,  while the various  lines indicate the
results  obtained from  three  different models  that  we describe  in
detail  below.  The (Poisson)  errors on  these model  predictions are
comparable to  those on the  P07 data and  are not shown for  clarity. 
Each  column and  row show,  respectively, the  results  for different
values of $n_{\rm crit}$ and $F_{\rm crit}$ as indicated.

We  start  our  investigation by  setting  $f_{d,{\rm  max}}=1$, which
implies that   the isophotal shape  of an  elliptical galaxy (disky or
boxy) is assumed  to  be independent of  the  amount of mass  accreted
since   the last major  merger.    Although  we  don't  consider  this
realistic, it is a useful starting point  for our investigation, as it
clearly separates the effects of $n_{\rm  crit}$ and $F_{\rm crit}$ on
the boxy fraction.  The results thus obtained from our semi-analytical
model with AGN feedback are  shown in Figure~\ref{fig:fboxy} as dotted
lines.   If we assign the isophotal  shapes of  ETGs depending only on
the progenitor  mass ratio $n$,  which  corresponds to setting $F_{\rm
  crit}=1$, we obtain the boxy fractions shown in the upper panels. In
agreement with   KB05 (see their  Figure  1)  this results  in  a boxy
fraction  that is   virtually  independent  of luminosity, in    clear
disagreement with the  data.  Note that, for  a given value of $n_{\rm
  crit}$,   our boxy fractions are   significantly higher  than in the
model of KB05.   For example, for $n_{\rm   crit}=2$ we obtain  a boxy
fraction of $\sim 0.6$, while KB05 find that $f_{\rm boxy} \sim 0.33$.
This mainly reflects the difference in the type  of merger trees used.
As discussed above, we use the merger  trees extracted from a $N$-body
simulation,  while KB05 use monte-carlo  merger trees based on the EPS
formalism.  It is well known that EPS merger  trees predict masses for
the most  massive progenitors that are too  large (e.g., Lacey \& Cole
1994; Somerville \etal 2000; van den  Bosch 2002; Benson, Kamionkowski
\&  Hassani  2005).  This implies  that  the number of  mergers with a
small progenitor mass ratio $n$ will  be too small, which explains the
difference between our results and  those of KB05.  Using cosmological
SPH simulations,  Maller \etal (2006) found   that the distribution of
merger  mass ratios  scales  as ${\rm d}N/{\rm  d}n \propto n^{-0.8}$.
This means that 60 percent  of all galaxy mergers with  $n < 3$ have a
progenitor  mass  ratio  $n  < 2$, in    excellent agreement  with our
results.

The  dotted lines  in the  remaining panels  of Figure~\ref{fig:fboxy}
show the  results obtained for  three different values of  the maximum
cold  gas  mass  fraction,  $F_{\rm  crit}=0.6$,  $0.3$,  and  $0.1$.  
Lowering $F_{\rm  crit}$ has a strong  impact on the  boxy fraction of
low-luminosity  ETGs,  while  leaving  that  of  bright  ETGs  largely
unaffected. As  we show  in \S\ref{sec:disc} this  mainly owes  to the
fact   that  $F_{\rm   cold}$  decreases   strongly   with  increasing
luminosity.  Consequently, by changing  $F_{\rm crit}$ we can tune the
slope  of the relation  between $f_{\rm  boxy}$ and  luminosity, while
$n_{\rm crit}$ mainly governs the absolute normalization.  We obtain a
good match to the P07 data for $n_{\rm crit}=2$ and $F_{\rm crit}=0.1$
(third panel  in lowest row).  This  implies that boxy  ETGs only form
out  of relatively  dry and  violent mergers,  in good  agreement with
numerical simulations.

\subsection{The influence of disk regrowth}
\label{sec:gasinfall}

The  analysis above,  however, does  not  consider the  impact of  the
growth of  a new disk around  the merger remnant. Since  this may turn
boxy systems into  disky systems, it can have  a significant impact on
the predicted $f_{\rm boxy}$.  We now take this effect into account by
setting  $f_{d,{\rm max}}$  to its  fiducial value  of $0.2$.

The solid  lines in Fig.~\ref{fig:fboxy} show the  boxy fractions thus
obtained.  A  comparison with  the dotted lines  shows that  the newly
formed disks  only cause a  significant decrease of $f_{\rm  boxy}$ at
the faint end. At the bright end, AGN feedback prevents the cooling of
hot gas, therewith significantly reducing the rate at which a new disk
can regrow\footnote{In the absence of cooling, the only way in which a
  galaxy can  (re)grow a disk  is via minor mergers.}.   However, when
$F_{\rm crit}=0.1$,  we obtain the same boxy  fractions for $f_{d,{\rm
    max}}=0.2$  as for  $f_{d,{\rm max}}=1$,  even at  the faint  end. 
This implies  that we obtain  the same conclusions as  above: matching
the data  of P07  requires $n_{\rm crit}  \simeq 2$ and  $F_{\rm crit}
\simeq 0.1$.   In other words,  our constraints on $n_{\rm  crit}$ and
$F_{\rm crit}$ are robust to exactly how much disk regrowth is allowed
before it reveals itself in the isophotes.

Why do faint  ETGs with $F_{\rm  cold}  < 0.1$ not regrow  significant
disks, while does with $F_{\rm cold} > 0.1$  do?  Note that during the
last major merger, the entire cold gas mass is converted into stars in
a starburst.   Therefore,  it is somewhat  puzzling that  the galaxy's
ability to regrow a disk depends on its cold gas  mass fraction at the
last  major merger.   As  it turns  out,  this  owes to  the fact that
progenitors with a  low cold gas mass fraction  are more  likely to be
satellite   galaxies.  Fig.~\ref{fig:cs} plots  the  fractions of ETGs
with last major mergers between a central galaxy and a satellite (C-S;
solid lines) and between two satellites (S-S; dashed lines). Note that
in our model, S-S mergers  occur whenever their dark matter sub-haloes
in the $N$-body  simulation  merge.  Results are   shown for all  ETGs
(left-hand  panel),  and  for    only  those ETGs  that   have $F_{\rm
  cold}<0.1$  (right-hand panel).   In our   fiducial model with   AGN
feedback (black lines) the  most massive ETGs almost  exclusively form
out   of C-S mergers.  Since a   satellite galaxy  can  never become a
central galaxy, this  is consistent with  the fact that virtually  all
massive  ETGs at $z=0$  are  central  galaxies  (in  massive  haloes).
Roughly 40 percent of   all low mass ETGs   have a last   major merger
between two  satellite galaxies.  However, when  we  only focus on low
mass ETGs with  $F_{\rm cold}<0.1$, we find that  $\sim 95$ percent of
their  last major mergers are  between  two satellite galaxies.  Since
the  $z=0$  descendents of S-S  mergers will  also be satellites, this
implies  that virtually all boxy ETGs  with $M_* \lta 2 \times 10^{10}
h^{-1}  \Msun$ are  satellite galaxies.  Furthermore,  since satellite
galaxies  do not have   a hot  gas  reservoir (at   least not  in  our
semi-analytical   model) they can  not regrow  a  new disk by cooling.
This  explains why  for  $F_{\rm  crit}=0.1$ the  boxy  fractions  are
independent of the value of $f_{d,{\rm max}}$.
\begin{figure*}
\centerline{\psfig{figure=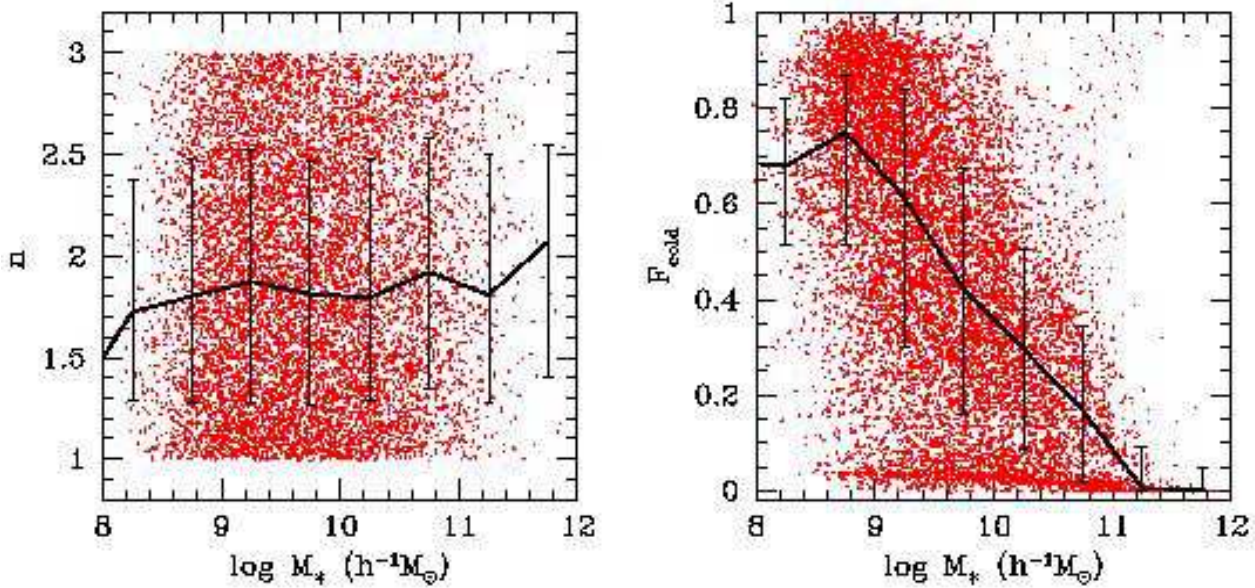,width=0.95\hdsize}}
\caption{The  progenitor mass ratio,   $n$, (left-hand panel) and  the
  cold gas  mass fraction at  the last  major merger, $F_{\rm  cold}$,
  (right-hand panel) as function of the $z=0$  stellar mass, $M_*$, of
  ETGs.   Solid  lines  with  errorbars  indicate the  median and  the
  $20^{\rm th}$  and $80^{\rm th}$  percentiles  of the distributions.
  While  the mass ratio of  the progenitors  of early-type galaxies is
  independent of its stellar  mass, $F_{\rm cold}$ decreases  strongly
  with increasing $M_*$.}
\label{fig:scatter}
\end{figure*}

\subsection{The role of AGN feedback}
\label{sec:AGN}

Our  semi-analytical  model   includes  ``radio-mode''  AGN  feedback,
similar to  that in  Croton \etal (2006),  in order to  suppresses the
cooling in massive haloes.  This  in turn shuts down star formation in
the central galaxies in these haloes, so that they become red.  In the
absence of AGN feedback, new  gas continues to cool making the central
galaxies in massive  haloes overly massive and too  blue (e.g., Benson
\etal 2003; Bower \etal 2006; Croton \etal 2006; Kang \etal 2006).  In
order to study its impact on $f_{\rm boxy}$ as function of luminosity,
we simply turn  off AGN feedback in our  model.  Although this results
in a semi-analytical  model that no longer fits  the galaxy luminosity
function at the bright end,  and results in a color-magnitude relation
with far too many bright,  blue galaxies, a comparison with the models
discussed above nicely  isolates the effects that are  directly due to
our prescription for AGN feedback.

The dashed lines in Figure~\ref{fig:fboxy} show the predictions of our
model  without  AGN  feedback  (and with  $f_{d,{\rm  max}}=0.2$).   A
comparison with our fiducial model (solid lines) shows that apparently
the AGN feedback  has no impact on $f_{\rm boxy}$  for faint ETGs with
$M_B  - 5\log  h \gta  -18$.   At the  bright end,  though, the  model
without AGN  feedback predicts  boxy fractions that  are significantly
lower (for  reasons that will be discussed  in \S\ref{sec:agnimpact}). 
Consequently,  the luminosity  dependence  of $f_{\rm  boxy}$ is  much
weaker than in  the fiducial case. The only model  that comes close to
matching  the data  of P07  is the  one with  $n_{\rm crit}  =  3$ and
$F_{\rm crit}  = 0.1$.  We emphasize,  though, that this  model is not
realistic.  In  addition to the  fact that this  semi-analytical model
does  not fit  the observed  luminosity function  and  color magnitude
relation, a value of $n_{\rm crit}=3$ is also very unlikely: numerical
simulations have clearly shown that mergers with a mass ratio near 1:3
almost  invariably result  in disky  remnants (e.g.,  Naab  \& Burkert
2003).

\section{Discussion}
\label{sec:disc}

\subsection{The Origin of the ETG Dichotomy}
\label{sec:dicotomy}

We  now examine  the physical  causes for  the various  scalings noted
above.  We start by investigating  why our fiducial model with $n_{\rm
  crit}=2$  and $F_{\rm  crit}=0.1$  is successful  in reproducing  the
luminosity dependence  of $f_{\rm boxy}$,  i.e., why it  predicts that
the boxy fraction increases with luminosity.  Given the method used to
assign  isophotal shapes to  the ETGs  in our  model, there  are three
possibilities: (i) brighter ETGs  have smaller progenitor mass ratios,
(ii)  brighter  ETGs  have  progenitors  with smaller  cold  gas  mass
fractions, or (iii) brighter ETGs  have less disk regrowth after their
last major merger.
\begin{figure*}
\centerline{\psfig{figure=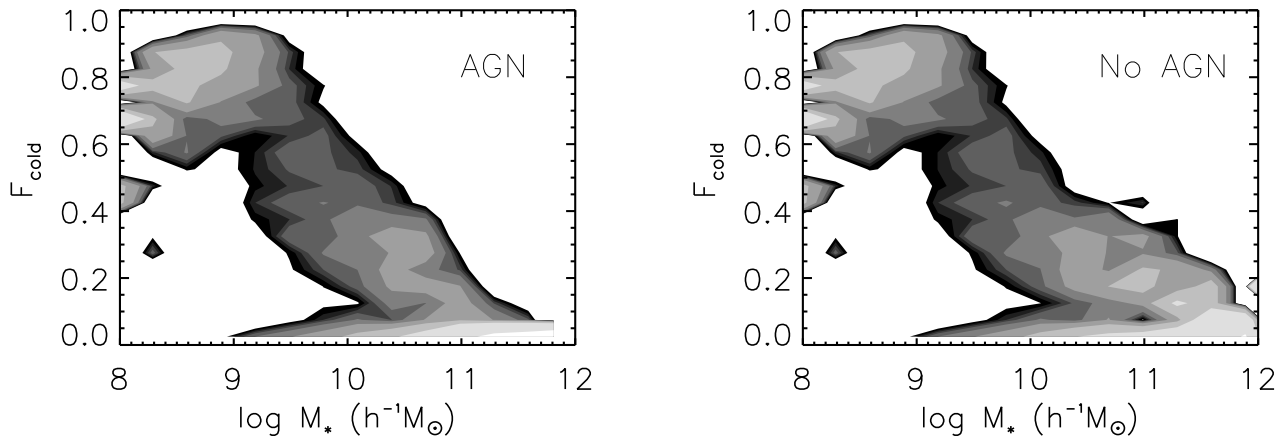,width=0.95\hdsize}}
\caption{Contour plots for the number  density of ETGs as function  of
  their  present  day stellar  mass, $M_*$,  and   their cold gas mass
  fraction  at the last  major  merger, $F_{\rm  cold}$.   Results are
  shown both for  our   fiducial model  with  AGN feedback  (left-hand
  panel), as  well as for  the model  without AGN feedback (right-hand
  panel).  In  both cases a  clear  bimodality is apparent:  ETGs with
  large and    low  masses  formed   out  of   dry and   wet  mergers,
  respectively.  Note  that this bimodality  is present independent of
  the presence of AGN feedback.}
\label{fig:fcold}
\end{figure*}

We can  exclude (i)  from the  fact that the  models that  ignore disk
regrowth (i.e., with $f_{d,{\rm max}}=1$) and that ignore the cold gas
mass  fractions (i.e., with  $F_{\rm crit}=1$)  predict that  the boxy
fraction is  roughly independent of luminosity (dotted  lines in upper
panels of Fig.~\ref{fig:fboxy}).   This suggests that the distribution
of $n$ is  roughly independent of the (present  day) luminosity of the
ETGs.    This   is   demonstrated    in   the   left-hand   panel   of
Fig.~\ref{fig:scatter}, were  we plot $n$  as function of,  $M_*$, the
stellar mass at $z=0$.  Each dot corresponds to an ETG in our fiducial
model, while  the solid  black line with  the errorbars  indicates the
median  and the  $20^{\rm th}$  and $80^{\rm  th}$ percentiles  of the
distribution:  clearly the  progenitor  mass ratio  is independent  of
$M_*$.

The boxy  fraction of our  best-fit model  with $n_{\rm  crit}=2$  and
$F_{\rm crit}=0.1$ is also independent of the regrowth of disks, which
is  evident from the   fact that the  models with  $f_{d,{\rm max}}=1$
(dotted lines)  and $f_{d,{\rm  max}}=0.2$ (solid lines)  predict boxy
fractions that are indistinguishable.  Therefore option (iii) can also
be excluded, and the luminosity  dependence of $f_{\rm boxy}$ thus has
to indicate that the  progenitors of more  luminous ETGs have a  lower
gas mass fraction.  That this is indeed the case  can be seen from the
right-hand panel of  Fig.~\ref{fig:scatter} which shows $F_{\rm cold}$
as function of  $M_*$.  Once  again,   the solid black  line with  the
errorbars indicates the median and the $20^{\rm th}$ and $80^{\rm th}$
percentiles of the  distribution.  Note that $F_{\rm cold}$  decreases
strongly  with increasing stellar   mass;  the most massive ETGs  form
almost exclusively from dry mergers with $F_{\rm cold} < 0.1$.

The  left-hand   panel  of  Fig.~\ref{fig:fcold}   shows  a  different
rendition of the relation  between $F_{\rm cold}$ and $M_*$.  Contours
indicate the number density,  $\phi(F_{\rm cold},M_*)$, of ETGs in the
$F_{\rm cold}$-$M_{*}$  plane, normalized by the total  number of ETGs
at each given $M_*$-bin, i.e.,
\begin{equation}
\int_0^1 \phi(F_{\rm cold},M_*) \, {\rm d}F_{\rm cold} = 1
\end{equation}
Note that  $\phi(F_{\rm cold},M_*)$ is clearly bimodal:  low mass ETGs
with $M_* \lta  3 \times 10^9 h^{-1} \Msun$  have high $F_{\rm cold}$,
while  the  progenitors  of  massive  ETGs  have  low  cold  gas  mass
fractions.  Clearly,  if wet  and dry mergers  produce disky  and boxy
ellipticals, respectively, this bimodality is directly responsible for
the ETG dichotomy.

What is  the physical  origin of this  bimodality?  It is  tempting to
expect that  it owes  to AGN  feedback.  After all,  in our  model AGN
feedback  is more  efficient  in more  massive  galaxies.  Since  more
massive  ETGs have more  massive progenitors,  one could  imagine that
their cold gas  mass fractions are lower because of  the AGN feedback. 
However, the right-hand panel  of Fig.~\ref{fig:fcold} shows that this
is not the case.  Here  we show $\phi(F_{\rm cold},M_*)$ for the model
without  AGN  feedback.  Somewhat  surprisingly,  this model  predicts
almost exactly  the same  bimodality as the  model with AGN  feedback. 
Their are  subtle differences, which  have a non-negligible  effect on
the    boxy   fractions    and    which   will    be   discussed    in
\S\ref{sec:agnimpact}  below.   However,   it  should  be  clear  from
Fig.~\ref{fig:fcold}  that  the  overall  bimodality  in  $\phi(F_{\rm
  cold},M_*)$ is not due to AGN feedback.

\begin{figure*}
\centerline{\psfig{figure=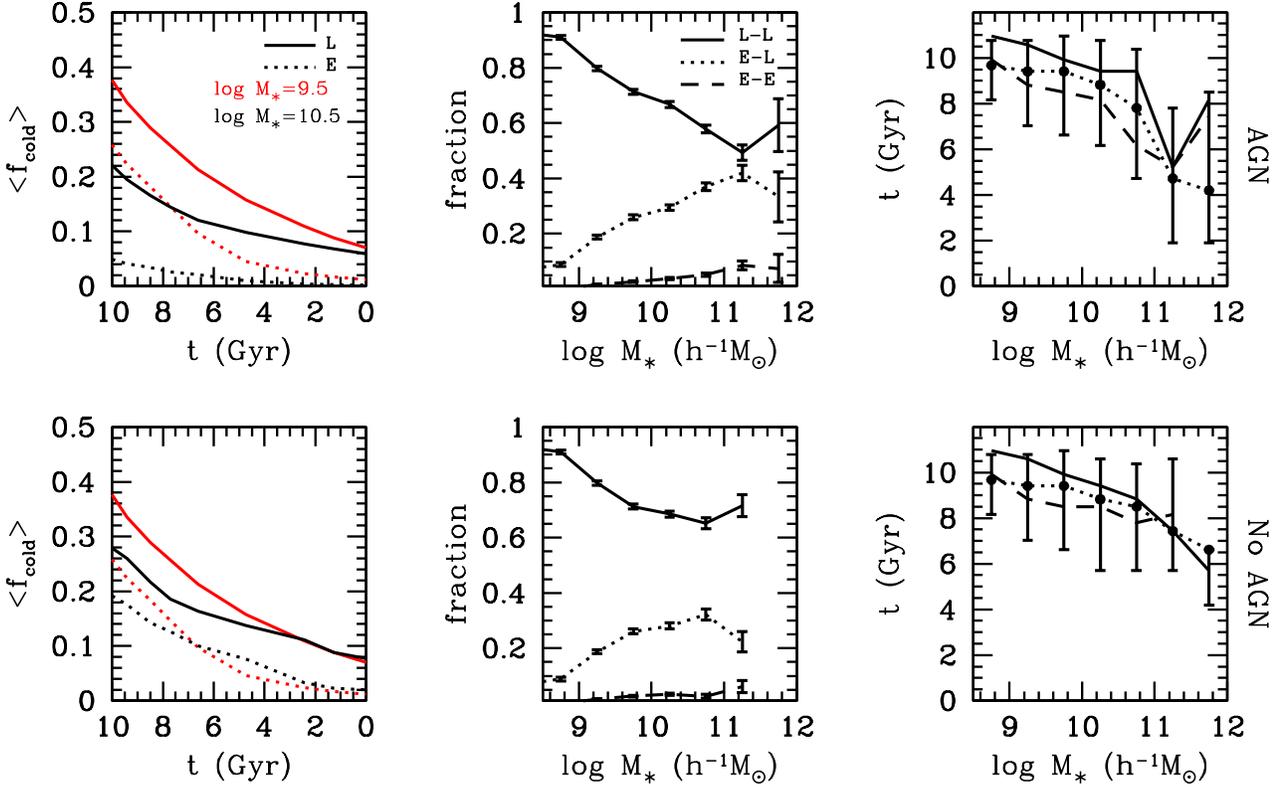,width=0.95\hdsize}}
\caption{Various statistics of our semi-analytical models. Upper and 
  lower  panels refer  to our  models with  and without  AGN feedback,
  respectively.   {\it Left-hand  panels:} The  average cold  gas mass
  fraction of  individual galaxies as  function of lookback time  $t$. 
  Results  are shown  for  galaxies of  two different  (instantaneous)
  stellar masses,  $M_* = 3\times  10^9 h^{-1} \Msun$ (red  lines) and
  $M_* =  3 \times  10^{10} h^{-1} \Msun$  (black lines), and  for two
  (instantaneous)  types: early-types  (dotted  lines) and  late-types
  (solid lines). {\it Middle panels:} The fractions of late-late (L-L;
  solid lines) , early-late  (E-L; dotted lines) and early-early (E-E;
  dashes lines) type mergers as  function of the $z=0$ stellar mass of
  the resulting  ETG.  {\it  Right-hand panels:} The  average lookback
  time to  the last major  merger of $z=0$  ETGs as function  of their
  $z=0$ stellar  mass.  Results are  shown separately for  L-L mergers
  (solid  line), E-L mergers  (dotted line),  and E-E  mergers (dashed
  line).  The  errorbars indicate the $20^{\rm th}$  and $80^{\rm th}$
  percentiles of the distribution of the E-L mergers.  For clarity, we
  do not  show these percentiles for  the L-L and  E-E mergers, though
  they are very similar. See text for detailed discussion.}
\label{fig:bimodal}
\end{figure*}

In  order  to explore  alternative  explanations  for the  bimodality,
Fig.~\ref{fig:bimodal} shows some relevant statistics. Upper and lower
panels   correspond to  the models   with   and without  AGN feedback,
respectively. Here we  focus on our  fiducial model with AGN feedback;
the  results for the model without  AGN  feedback will be discussed in
\S\ref{sec:agnimpact}.   The upper left-hand  panel shows  the average
cold  gas mass fraction of individual  galaxies, $\langle f_{\rm cold}
\rangle$, as function of lookback time.  Note that here we use $f_{\rm
  cold}$ to distinguish it  from  $F_{\rm cold}$, which indicates  the
cold  gas mass fraction of the  combined progenitors taking  part in a
major  merger, as defined in  eq.~(\ref{fcolddef}).  Results are shown
for galaxies of two  different (instantaneous) stellar masses, $M_*  =
3\times 10^9 h^{-1}  \Msun$ (red lines) and  $M_*  = 3 \times  10^{10}
h^{-1}  \Msun$   (black lines),  and   for two  (instantaneous) types:
early-types  (dotted lines)  and late-types  (solid lines).  Following
N06, here we  define  early-types  as  systems with  a  bulge-to-total
stellar mass ratio of  0.6 or larger; contrary  to our $z=0$ selection
criteria  described in \S\ref{sec:model}, we   do not include a  color
selection, simply because  the overall color  of the galaxy population
evolves as function of time.  First of all,  note that $\langle f_{\rm
  cold} \rangle$ of  galaxies of given  type and  given mass decreases
with increasing  time (i.e., with  decreasing lookback time).  This is
simply due to the consumption by star formation.  Secondly, at a given
time, early-type galaxies have lower gas mass fractions than late-type
galaxies.  This mainly owes to the fact  that at a major merger, which
creates an  early-type,  all the available  cold gas  is consumed in a
starburst.  Consequently, each early-type starts its life with $f_{\rm
  cold}=0$.  Finally,  massive galaxies have  lower gas mass fractions
than their less massive counterparts.  This owes to the fact that more
massive galaxies live, on average, in  more massive haloes, which tend
to  form  (not assemble!)  earlier   thus  allowing star  formation to
commence at an earlier epoch (see Neistein  \etal 2006).  In addition,
the star formation efficiency  used  in the semi-analytical  model  is
proportional to the  mass of the  cold gas times $M_{\rm vir}^{0.73}$.
As  discussed   in K05, this  scaling with   the halo virial   mass is
required    in order to  match  the    observed $\langle f_{\rm  cold}
\rangle(M_*)$ at $z=0$ (see also Cole \etal 1994, 2000; De Lucia \etal
2004).

The middle panel in the upper row of Fig.~\ref{fig:bimodal} shows what
kind  of  galaxy types are  involved   in the  last major  mergers  of
present-day ETGs.  Solid, dotted and  dashed curves show the fractions
of L-L, E-L and E-E mergers, where `L' and `E' refer to late-types and
early-types, respectively.  As above, these types  are based solely on
the bulge-to-total mass  ratio of the galaxy and  not on its color. In
our semi-analytical model,  the  lowest mass ETGs  almost  exclusively
form  via L-L mergers.  With increasing  $M_*$,   however, there is  a
pronounced decrease of  the fraction of L-L  mergers, which are mainly
replaced by E-L  mergers.  The fraction  of E-E mergers increases very
weakly with increasing stellar mass    but never exceeds 10   percent.
Thus, although boxy ellipticals form out of dry mergers, these are not
necessarily mergers between early-type  galaxies.  In fact, our  model
predicts that  the vast majority  of all dry  mergers involve at least
one late-type galaxy (though with a low cold gas mass fraction).  This
is in  good agreement with the SPH  simulation of Maller \etal (2006),
who also  find that E-E mergers are   fairly rare.  However,  it is in
stark contrast to the predictions of the semi-analytical model of N06,
how  find that more  than  50 percent of    the last major mergers  of
massive ellipticals are E-E mergers.   We suspect that the main reason
for  this strong discrepancy  is the fact  that N06  used merger trees
based on the EPS formalism.

Finally, the  upper right-hand  panel of Fig.~\ref{fig:bimodal}  plots
the average lookback time to the last major merger of ETGs as function
of their  present day stellar  mass.  Results are shown separately for
L-L mergers (solid line),  E-L mergers (dotted  line), and E-E mergers
(dashed line).  Clearly,  more massive ETGs  assemble  later (at lower
lookback  times).  This mainly   owes to the  fact that   more massive
galaxies live in more massive haloes,  which themselves assemble later
(cf.  Lacey \&  Cole 1993; Wechsler  \etal 2002;  van den  Bosch 2002;
Neistein \etal  2006; De Lucia \etal  2006).  In addition, it is clear
that at fixed stellar mass, E-E mergers  occur later than L-L mergers,
with  E-L  mergers in  between.   This difference,  however,  is small
compared to the scatter.
  
If we combine all this  information, we infer that the bimodality in
$\phi(F_{\rm   cold},M_*)$ owes to the following three facts:
\begin{itemize}
  
\item More  massive ETGs have  more massive progenitors  (this follows
  from the  fact that $n$ is  independent of $M_*$). Since  at a given
  time more massive galaxies of a  given type have lower cold gas mass
  fractions, $\langle F_{\rm  cold} \rangle$ decreases with increasing
  $M_*$.
  
\item More massive ETGs  assemble later (at lower redshifts). Galaxies
  of  given mass  and  given  type have  lower  $\langle f_{\rm  cold}
  \rangle$  at  later  times.   Consequently,  $\langle  F_{\rm  cold}
  \rangle$ decreases with increasing $M_*$.
  
\item  More  massive  ETGs   have  a  larger  fraction  of  early-type
  progenitors.   ETGs of  a  given mass  have  a lower  cold gas  mass
  fraction than late type galaxies of  the same mass, at any redshift. 
  In  addition, E-L mergers  occur at  later times  than L-L  mergers. 
  Both these effects also contribute  to the fact that $\langle F_{\rm
    cold} \rangle$ decreases with increasing $M_*$.

\end{itemize}

\subsection{Is AGN feedback relevant?}
\label{sec:agnimpact}

A comparison  of the upper and lower  panels in Fig.~\ref{fig:bimodal}
shows that the three effects  mentioned above, and thus the bimodality
in $\phi(F_{\rm cold},M_*)$, are present independent of whether or not
the model  includes feedback from  active galactic nuclei.   There are
only two small differences: without  AGN feedback massive ETGs (i) are
more  likely  to result  from  L-L mergers,  and  (ii)  have a  higher
$\langle  f_{\rm  cold} \rangle$  (cf.   black  dotted  curves in  the
left-hand  panels of  Fig.~\ref{fig:bimodal}).   Both effects  reflect
that  AGN feedback  prevents the  cooling  of hot  gas around  massive
galaxies, therewith  removing an important channel for  building a new
disk.  As is evident from Fig.~\ref{fig:fcold}, these two effects only
have  a very mild  impact on  $\phi(F_{\rm cold},M_*)$.   We therefore
conclude that {\it the bimodality of ETGs is not due to AGN feedback.}
\begin{figure}
\centerline{\psfig{figure=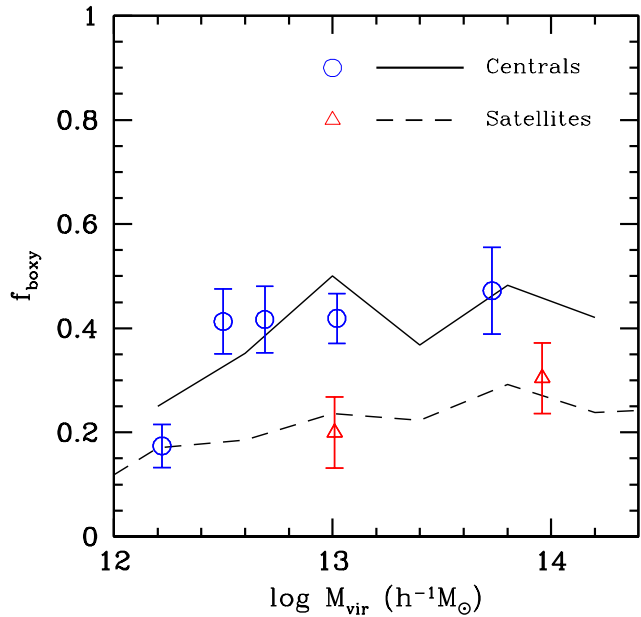,width=\hssize}}
\caption{The boxy fraction of  ETGs as function  of halo (group) mass.
  Red triangles (for satellite galaxies) and blue circles (for central
  galaxies) are  taken from  P07,  and have   been obtained  using the
  galaxy  group catalogue of Weinmann  \etal (2006).  Dashed and solid
  lines correspond to the predictions from our fiducial model.}
\label{fig:fbox-mass}
\end{figure}

This  does not  imply, however,  that AGN  feedback does  not  have an
impact on the boxy fractions. As is evident from Fig.~\ref{fig:fboxy},
the  models with and  without AGN  feedback clearly  predict different
$f_{\rm boxy}$ at  the bright end.  To understand  the origin of these
differences,  first  focus  on  Fig.~\ref{fig:fcold}.   Although  both
panels look very similar, upon  closer examination one can notice that
at $M_*  \gta 10^{11}  h^{-1} \Msun$ the  number density of  ETGs with
$0.1 \lta F_{\rm cold} \lta 0.25$ is significantly larger in the model
without AGN feedback. In the model with AGN feedback these systems all
have $F_{\rm  cold} <  0.1$. This explains  why the model  without AGN
feedback  predicts a  lower  boxy fraction  for  bright galaxies  when
$F_{\rm crit}=0.1$. However, this  does not explain why $f_{\rm boxy}$
is also different when $F_{\rm  crit} \geq 0.3$.  After all, for those
models  it should not  matter whether  $F_{\rm cold}=0.05$  or $F_{\rm
  cold}=0.25$,  for example.   It turns  out that  in these  cases the
differences between the  models with and without AGN  feedback are due
to  the regrowth  of a  new disk;  since AGN  feedback  suppresses the
cooling of hot gas around massive galaxies, it strongly suppresses the
regrowth of  a new disk, thus  resulting in higher  boxy fractions. 

Note however, that in ETGs with $F_{\rm cold} < 0.1$, disk regrowth is
always negligible. In the presence of  AGN feedback this is due to the
suppression  of cooling  in massive  haloes.   In the  absence of  AGN
feedback it owes  to the fact that only a very  small fraction of ETGs
are  central galaxies. As  can be  seen from  the right-hand  panel of
Fig.~\ref{fig:cs}, more  than 90 percent  of the ETGs have  last major
mergers  between  two  satellite  galaxies  (with  AGN  feedback  this
fraction is smaller than 20  percent). Since satellite galaxies do not
have hot gas reservoirs, no  significant disks can regrow around these
systems.

\subsection{Environment dependence}
\label{sec:environ}

Using the SDSS galaxy group  catalogue of Weinmann \etal (2006), which
has been  constructed using the  halo-based group finder  developed by
Yang  \etal (2005), P07  investigated how  $f_{\rm boxy}$  scales with
group  mass.   They also  split  their  sample  in `central'  galaxies
(defined as  the brightest group members) and  `satellites'.  The open
circles and  triangles in Fig.~\ref{fig:fbox-mass}  show their results
for centrals  and satellites,  respectively.  Although there  are only
two data points for the  satellites, it is clear that central galaxies
are more likely  to be boxy than a satellite galaxy  in a group (halo)
of the same mass.

We now investigate whether  our fiducial semi-analytic model that fits
the luminosity  dependence of  the boxy fraction  (i.e., the  one with
$n_{\rm  crit}=2$ and  $F_{\rm  crit}=0.1$) can  also reproduce  these
trends.   The model predictions  for the  centrals and  satellites are
shown   in  Fig.~\ref{fig:fbox-mass}  as   solid  and   dashed  lines,
respectively. Here  we have associated  the halo virial mass  with the
group mass,  and an ETG is  said to be a  central galaxy if  it is the
brightest  galaxy in its  halo.  The  model accurately  reproduces the
boxy fraction  of both central and satellite  galaxies. In particular,
it  reproduces the  fact that  $f_{\rm boxy}$  of central  galaxies is
higher than that of satellites in groups (haloes) of the same mass.

As shown in P07, the boxy fraction as function of group mass, for both
centrals   and   satellites,   is   perfectly  consistent   with   the
null-hypothesis that the isophotal shape of an ETG depends only on its
luminosity; the  fact that centrals  have a higher boxy  fraction than
satellites  in  the same  group,  simply owes  to  the  fact that  the
centrals  are brighter.   Also, the  increase of  $f_{\rm  boxy}$ with
increasing group  mass simply reflects  that more massive  haloes host
brighter galaxies.  It  therefore may not come as  a surprise that our
semi-analytical model  that fits the luminosity  dependence of $f_{\rm
  boxy}$   also   fits   the   group  mass   dependencies   shown   in
Fig.~\ref{fig:fbox-mass}. It does mean,  though, that in our model the
merger histories of ETGs of  a given luminosity do not strongly depend
on the halo mass in which the galaxy resides.
\begin{figure*}
\centerline{\psfig{figure=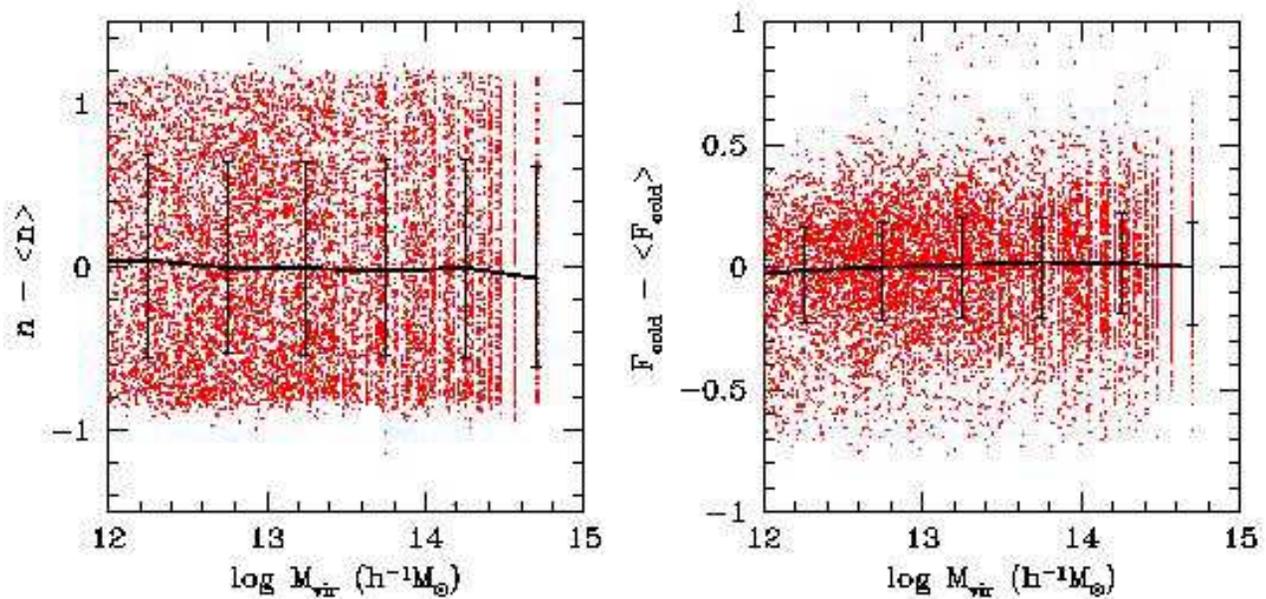,width=0.95\hdsize}}
\caption{The  residuals of the relations   between $n$ and $M_*$ (left
  panel)  and $F_{\rm cold}$  and $M_*$ (right  panel) as functions of
  the virial mass of the halo in  which the ETGs  reside at $z=0$.  As
  in  Fig.~\ref{fig:scatter} the solid   lines with errorbars indicate
  the mean and the $20^{\rm th}$ and $80^{\rm  th}$ percentiles of the
  distributions.  These show that  after  one corrects for the stellar
  mass dependence, the properties  of the last  major mergers  of ETGs
  are independent of their halo mass.}
\label{fig:residual}
\end{figure*}

To  test  this we proceed   as follows. For  each ETG  in our model we
compute $\langle n \rangle$ and $\langle  F_{\rm cold} \rangle$, where
the  average is over all  ETGs with stellar masses  similar to that of
the galaxy  in question.   Fig.~\ref{fig:residual} plots the residuals
$n   - \langle n \rangle$  and  $F_{\rm cold}  -  \langle F_{\rm cold}
\rangle$ as function of the virial mass, $M_{\rm vir}$, of the halo in
which they reside.  This clearly shows that after one corrects for the
stellar mass  dependence, the properties of the  last major  merger of
ETGs are indeed independent of  their halo mass\footnote{The fact that
  the distribution of the progenitor mass  ratio $n$ is independent of
  halo mass   was  also found by   KB05}.    This provides theoretical
support for the   conclusion  of   P07  that  the stellar  mass    (or
luminosity) of an ETG is the main parameter that determines whether it
will be disky or boxy.

\subsection{The Origin of Cusps and Cores}
\label{sec:core}

As discussed in  \S\ref{sec:intro}, the dichotomy of ETGs  is not only
restricted to  their isophotal shapes.  One other  important aspect of
the dichotomy regards the  central density distribution of ETGs; while
disky  systems typically  have  cuspy profiles,  the  bright and  boxy
ellipticals generally reveal density  profiles with a pronounced core. 
Here we briefly  discuss how the formation of cusps  and cores fits in
the picture sketched above.

In the paradigm adopted here, low  luminosity ETGs form mainly via wet
mergers.  Due to the fluctuating potential of the merging galaxies and
the  onset  of bar instabilities,  the gas  experiences strong torques
which causes  a significant fraction  of  the gas to sink  towards the
center of the  potential well  where  it undergoes a starburst  (e.g.,
Shlosman, Frank \&  Begelman 1989; Barnes \&  Hernquist 1991; Mihos \&
Hernquist 1996).    Detailed  hydrodynamical  simulations  of gas-rich
mergers (e.g., Springel \& Hernquist 2005; Cox  \etal 2006b) result in
the formation  of remnants with  surface brightness profiles  that are
reminiscent of cuspy   ETGs  (John Kormendy, private   communication).
Hence, cusps seem  a natural by-product  of the dissipative  processes
associated with a wet merger.

\begin{figure}
\centerline{\psfig{figure=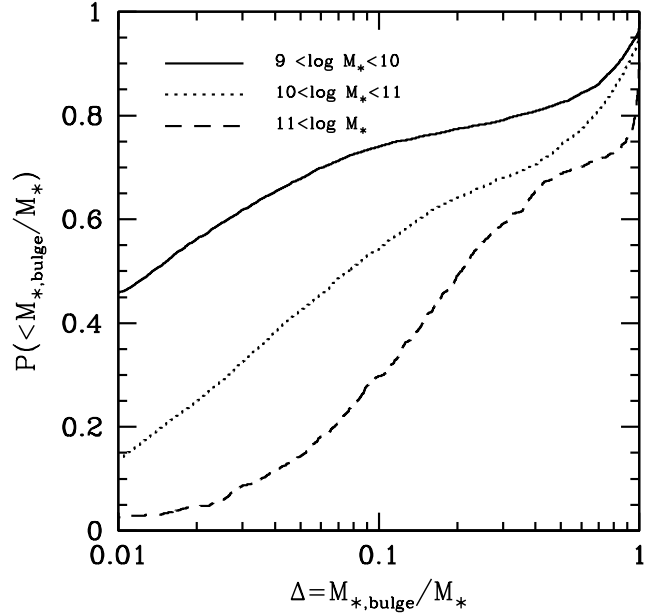,width=\hssize}}
\caption{The cumulative  probability that a progenitor  of a $z=0$ ETG
  has  a  stellar  bulge-to-total  mass  ratio  less  then  $M_{*,{\rm
      bulge}}/M_*$. Results  are shown for the progenitors  of ETGs in
  three  stellar mass  ranges,  as indicated  (masses  are in  $h^{-1}
  \Msun$).   Note  that the  progenitors  of  more  massive ETGs  have
  significantly higher  $M_{*,{\rm bulge}}/M_*$.  As  discussed in the
  text, this may  help to explain why low mass  ETGs have cusps, while
  massive ones have cores.}
\label{fig:btod}
\end{figure}

Boxy ETGs, however, are  thought to form  via dry  mergers. As  can be
seen from Fig.~\ref{fig:bimodal} roughly  $35$ percent of all  massive
ETGs originate from  a last major  merger that involves an  early-type
progenitor.  If this progenitor contains a cusp, this will survive the
merger,   as  most clearly  demonstrated  by Dehnen  (2005).  The only
mergers that are believed to result directly in a remnant with a core,
are  mergers between  pure   stellar disks with  a  negligible $f_{\rm
  cold}$  (e.g.,  Cox  \etal   2006a).  Fig.~\ref{fig:btod} shows  the
cumulative distributions of  the bulge-to-total stellar mass ratios of
the progenitors  of present day ETGs.  Results  are shown for  ETGs in
three mass ranges, as indicated.  The probability that a progenitor of
a  massive ETGs (with $M_* >  10^{11}  h^{-1} \Msun$) has a negligible
bulge    component ($M_{*,{\rm bulge}} < 0.01    M_*$) is only about 3
percent.  Hence, we expect  that only about 1  out of every 1000 major
mergers that result in a massive ETG will have  a remnant with a core.
And  this is most likely  an overestimate, since we  did  not take the
cold gas mass fractions  into consideration.  Since the  cusp accounts
for only about   one percent of  the total  stellar mass (e.g.,  Faber
\etal 1997; Milosavljevi\'c \etal 2002), cold gas  mass fractions of a
few percent  are probably enough  to  create a  cusp via dissipational
processes.

Therefore, an  additional mechanism is  required in order to  create a
core (i.e., destroy a cusp).  Arguably the most promising mechanism is
the orbital  decay of a  supermassive black hole (SMBH)  binary, which
can scour  a core by exchanging  angular momentum with  the cusp stars
(e.g., Begelman \etal 1980;  Ebisuzaki \etal 1991; Quinlan 1996; Faber
\etal  1997;  Milosavljevi\'c   \etal  2002;  Merritt  2006a).   Since
virtually all  spheroids contain a SMBH  at their center,  with a mass
that  is tightly  correlated  with  the mass  of  the spheroid  (e.g.,
Kormendy \& Richstone 1995;  Ferrarese \& Merritt 2000; Gebhardt \etal
2000; Marconi  \& Hunt  2003; H\"aring \&  Rix 2004), it  is generally
expected that  such binaries  are common in  merger remnants  (but see
below).

While offering an attractive explanation  for the presence of cores in
massive,  boxy ETGs,  this  picture simultaneously  poses a  potential
problem for  the presence of cusps  in disky ETGs.  After  all, if the
progenitors of  disky ETGs also  harbor SMBHs, the same  process could
create a core  in these systems as well.  There  are two possible ways
out of this paradox:  (i) low mass ETGs do not form  a SMBH binary, or
(ii) a cusp is regenerated after  the two SMBHs have coalesced. We now
discuss these two options in turn.

In  order for a  SMBH binary  to form,  dynamical friction  must first
deliver the  two SMBHs from the  two progenitors to the  center of the
newly formed merger  remnant.  This process will only  be efficient if
the spheroidal hosts of the SMBHs are sufficiently massive. Consider a
(small) bulge  that was  part of a  late-type progenitor which  is now
orbiting the  remnant of its  merger with another galaxy.   Assume for
simplicity  that both  the bulge  and  the merger  remnant are  purely
stellar  singular  isothermal  spheres  ($\rho \propto  r^{-2}$)  with
velocity dispersions equal to $\sigma_b$ and $\sigma_g$, respectively.
Then, assuming that the bulge is  on a circular orbit, with an initial
radius $r_i$, Chandrasekhar's (1943)  formula gives an infall time for
the bulge of
\begin{equation}
t_{\rm infall} \approx 3.3 {r_i \over \sigma_g} 
\left(\sigma_g \over \sigma_b\right)^3 
\approx 4.7 \times 10^8 \yr \left(M_g \over M_b \right)
\end{equation}
(Merritt 2006b). Here we  have used that $r_i/\sigma_g \simeq \sqrt{2}
\,  t_{\rm cross}$  with  $t_{\rm  cross} \sim  10^8  \yr$ the  galaxy
crossing time.  If the galaxy is  the remnant of an equal mass merger,
so that  $M_b \sim (\Delta/2)  M_g$, with $\Delta$  the bulge-to-total
stellar mass ratio  of the late-type progenitor, we  find that $t_{\rm
  infall}$ is equal to the  Hubble time ($1.3 \times 10^{10} \yr$) for
$\Delta \simeq  0.07$. As can be seen  from Fig.~\ref{fig:btod}, about
70  percent of the  low mass  ETGs (with  $10^9 h^{-1}  \Msun <  M_* <
10^{10} h^{-1}  \Msun$) have  at least one  progenitor with  a stellar
bulge-to-total mass ratio $\Delta  < 0.07$.  Therefore, we expect that
a similar fraction will form without  a SMBH binary, and thus will not
form a  core. For  comparison, for massive  ETGs (with $M_*  > 10^{11}
h^{-1} \Msun$)  only about 20 percent  of the progenitors  will have a
sufficiently small bulge to prevent the formation of a SMBH binary.

An alternative explanation for the presence of  cusps in low mass ETGs
is that the cusp is regenerated by  star formation from gas present at
the  last major merger. However, as  emphasized by Faber \etal (1997),
this results in a serious timing problem, as it  requires that the new
stars    must form  after  the  SMBH   binary  has coalesced.  Another
potential problem with this picture, is that the cusp would be younger
than  the main body  of the ETG which  may  lead to observable effects
(i.e., cusp could be bluer  than main body). However,  in light of the
results  presented here, we believe that  neither of  these two issues
causes a serious  problem.  First of  all, the cold gas mass fractions
involved with the last major merger, and hence  the mass fraction that
is turned  into stars in the  resulting starburst, is extremely large:
$\langle F_{\rm cold}  \rangle \sim  0.8$ (see  Fig.~\ref{fig:fcold}).
As mentioned above, a significant fraction of  this gas is transported
to the center, where it will function  as an important energy sink for
the SMBH binary,   greatly speeding up  its coalescence  (Escala \etal
2004, 2005) and therewith reducing the timing problem mentioned above.
In fact, the  gas may well  be the dominant  energy sink, so  that the
pre-existing  cusps of the progenitors  are only mildly affected.  But
even  if the cusps were destroyed,  there clearly should be enough gas
left  to  build a   new  cusp.  In   fact,  if, as  envisioned in  our
semi-analytical model, all the cold   gas present  at the last   major
merger is consumed in a starburst,  a very significant fraction of the
stars in the main body  would also  be  formed in this starburst  (not
only  the  cusp).  This would  help   to diminish potential population
differences  between  the cusp  and  the main  body   of the  ETG.  In
addition,   as   can   be    seen from   the    right-hand  panels  of
Fig.~\ref{fig:bimodal}, the  last major merger  of low luminosity ETGs
occurred  on average $\sim 9.5\;$Gyr  ago.   Hence, the stars made  in
this burst are not  easily distinguished observationally from the ones
that were already present before the last major merger.

To summarize, our semi-analytical  model predicts that the progenitors
of ETGs  have cold gas  mass fractions and bulge-to-total  mass ratios
that offer a relatively natural explanation for the observed dichotomy
between cusps and cores.

\section{Conclusions}
\label{sec:concl}

Using a  semi-analytical model for  galaxy formation, combined  with a
large  $N$-body simulation,  we have  investigated the  origin  of the
dichotomy among ETGs.  In order to assign isophotal shapes to the ETGs
in our model we  use three criteria: an ETG is said  to be boxy if (i)
the progenitor  mass ratio  at the  last major merger  is $n  < n_{\rm
  crit}$, (ii) the total cold gas  mass fraction of the sum of the two
progenitors at the last major merger is $F_{\rm cold} < F_{\rm crit}$,
and (iii) after its last major merger the ETG is not allowed to regrow
a new  disk with a stellar mass  that exceeds 20 percent  of the total
stellar mass. 

In agreement with KB05, we find that we can not reproduce the observed
luminosity (or, equivalently,   stellar  mass) dependence  of  $f_{\rm
  boxy}$  if we assign  isophotal shapes based  only on the progenitor
mass ratio.  This  owes to the  fact that the  distribution of $n$  is
virtually independent of the stellar mass, $M_*$, of the ETG at $z=0$.
Rather,  to obtain  a  boxy  fraction that  increases  with increasing
luminosity one also  needs to consider  the cold gas mass  fraction at
the last  major merger. In fact,  we can accurately  match the data of
P07 with $n_{\rm crit}=2$  and $F_{\rm crit}=0.1$.  This  implies that
boxy galaxies originate from  relatively violent and dry  mergers with
roughly equal mass progenitors and with less than 10 percent cold gas,
in good agreement with numerical  simulations (e.g., Naab \etal 2006a;
Cox \etal 2006a).  Our model also  nicely reproduces the observed boxy
fraction  as function of   halo mass, for   both central galaxies  and
satellites.  We  have  demonstrated that this  owes to  the  fact that
after one corrects for the  stellar mass dependence, the properties of
the last major merger of ETGs are independent of their halo mass. This
provides theoretical support for  the   conjecture  of P07 that    the
stellar mass  (or  luminosity) of an   ETG is the  main parameter that
determines whether it will be disky or boxy.

Our  model   predicts   a number  density   distribution, $\phi(F_{\rm
  cold},M_*)$, of  ETGs in the   $F_{\rm cold}$-$M_{*}$ plane  that is
clearly bimodal: low mass ETGs  with  $M_* \lta  3 \times 10^9  h^{-1}
\Msun$ have high $F_{\rm cold}$, while the progenitors of massive ETGs
have low cold gas  mass fractions.  Clearly,   if wet and dry  mergers
produce disky and  boxy ellipticals, respectively, this  bimodality is
directly  responsible  for  the  ETG   dichotomy.  Contrary to   naive
expectations,  we   find that this  bimodality   is independent of the
inclusion  of  AGN feedback in the  model.   Although AGN  feedback is
essential for regulating the luminosities and  colors of the brightest
galaxies (which   end up  as  ETGs  with  AGN  feedback, but  as  blue
disk-dominated systems without AGN  feedback), it does not explain the
bimodality among  ETGs.   Rather, this bimodality  is due  to the fact
that more    massive ETGs  (i)  have  more  massive  progenitors, (ii)
assemble  later, and  (iii)  have   a  larger fraction of   early-type
progenitors.   Each of these  three trends  causes  the cold gas  mass
fraction of the progenitors of more massive ETGs to be lower, and thus
its last major merger to be dryer.  In conclusion, the dichotomy among
ETGs has a very  natural explanation within the hierarchical framework
of structure formation and does not require AGN feedback.

We also  examined the morphological  properties of the  progenitors of
present day ETGs  (at the epoch of the  last major merger). Indicating
early- and late-type galaxies with  `E' and `L', respectively, we find
that the  lowest mass  ETGs almost exclusively  form via L-L  mergers. 
With increasing $M_*$, however, there  is a pronounced decrease of the
fraction of  L-L mergers,  which are mainly  replaced by E-L  mergers. 
The E-E  mergers, however, never  contribute more than 10  percent, in
good agreement with the SPH simulations of Maller \etal (2006).  Thus,
although boxy ellipticals  form out of dry mergers,  these only rarely
involve two early-type systems.

Since satellite  galaxies do not have a  hot corona from which new gas
cools  down, they  typically have  lower cold  gas mass fractions than
central galaxies  of the same  mass.   Consequently, dry  mergers  are
preferentially mergers between two satellite galaxies.  In fact, since
a satellite galaxy can not become a central galaxy, our model predicts
that more  than 95 percent of  all boxy ETGs with  $M_*  \lta 2 \times
10^{10} h^{-1} \Msun$ are satellites.

We also find that the  progenitors of less massive ETGs typically have
lower bulge-to-total mass ratios.  In  fact, for ETGs with present day
stellar masses in the range $10^9  h^{-1} \Msun < M_* < 10^{10} h^{-1}
\Msun$ we find  that almost half of the progenitors  at the last major
merger have bulges that do not contribute more than one percent to the
total  stellar mass.   This may  have important  implications  for the
observed dichotomy between cusps and cores in ETGs. Cores are believed
to form via the scouring effect of a SMBH binary, that arises when the
SMBHs  associated with  the  spheroidal components  of the  progenitor
galaxies form  a bound pair. This  requires both spheroids  to sink to
the center of  the potential well of the  merger remnant via dynamical
friction.   However, if  the time  scale for  this infall  exceeds the
Hubble time, no SMBH binary will form, thus preventing the creation of
a core.   Using our prediction  for the bulge-to-total mass  ratios of
progenitor galaxies,  and a  simple estimate based  on Chandrasekhar's
dynamical friction  formula, we have estimated that  $\sim 70$ percent
of low mass ETGs in the aforementioned mass range will not form a SMBH
binary.   For massive  ETGs with  $M_*  > 10^{11}  h^{-1} \Msun$  this
fraction is only $\sim 20$ percent.   This may help to explain why low
mass ETGs have steep cusps, while massive ETGs have cores.

Finally, in  those low mass  systems that  do form a  SMBH binary, the
large cold gas mass fraction at its last major merger ($\langle F_{\rm
  cold} \rangle \simeq  0.8$) provides more  than enough raw  material
for the regeneration of a new cusp.  In addition,  a large fraction of
the cold gas will sink to the center due  to angular momentum transfer
where   it  will function  as an  important  energy sink  for the SMBH
binary.  As shown   by Escala \etal  (2004, 2005),  this can  cause  a
tremendous acceleration  of  the  coalescence of the    SMBHs, largely
removing the timing problem interjected by Faber \etal (1997).

\section{Acknowledgements}

We are grateful  to Eric Bell, Eric Emsellem,  John Kormendy, Thorsten
Naab, Hans-Walter Rix,  and the entire Galaxies-Cosmology-Theory group
at the MPIA for enlightening discussions.



\label{lastpage}
\end{document}